# CONCEPTUAL FRAMEWORK OF REDUNDANT LINK AGGREGATION

Rafiullah Khan[1] andShaukat Ali[2]

[1]Institute of Business and Management Science,TheUniversity of Agriculture Peshawar, Pakistan
rafiyz@gmail.com
[2]FG Degree College Peshawar Cantt, KPK, Pakistan Peshawar, Pakistan
shoonikhan@gmail.com

## *ABSTRACT*

*This is era of information blast. A huge quantity of information is pouring in from various sources. The revolutionary advancement of Information and Communication technologies bring the world close together.A pile of information in different formats is just a click away. Which motivate the organizations to get more internet bandwidth to consume and publish theinformationoverexploding cloudof Internet. The standard router redundancyprotocolis used to handle backup link showever it cannot aggregate them.Whereas thelink standard aggregation protocol can aggregate the link but it support only Ethernet technology.In this researchpaper a concept of Redundant Link Aggregation (RLA)is proposed. RLA can aggregate and handle backup links with main links regardless of carriertechnology. Furthermore a dataforwardingmechanism Odd Load Balancing (OLB) is also proposed for RLA scheme. For the sake of performance evaluation, Redundant Link Aggregation (RLA) is compared with Virtual Router Redundancy Protocol (VRRP). The simulation result reveals that Redundant Link Aggregation (RLA) can cover the bandwidth demand of the network in peak hours by consuming backup links as well which with Virtual Router Redundancy Protocol (VRRP)cannot.It is further noted thatOdd Load Balancing (OLB) feature can be used to save the cost in terms of money per annum.*

## 1. INTRODUCTION

The recent advancement in Information and Communication Technologies has shrunk the World. A new term of Global Village is now adopted, thanks to the instant information sharing property of electronic media. The ease in data publishing facilitate the organizations to unleash their data, due to which Internet became the mammoth of information. In February 2013 “WWW Size” website reported 13.02 billion indexed webpages1. Today’s web is completely different from early web. Early web was created to transfer the textual data; however today almost all data publishers publishes multimedia and data of other services as well. The access and exploitation of this data requires high bandwidth connectivity with Internet. The need of higher bandwidth connectivity is also compulsory for publishing the multimedia data and Voice or Video communication.

[1] http://www.worldwidewebsize.com

The need of more bandwidth is not new a thing. Many organizations of developed countries predict the future need, and start research for increasing bandwidth at different levels of network. In this regard notable contribution was development of Integrated Services Digital Network (ISDN) in 1988 [1]. The idea behind the ISDN was to combine different number of telephone lines in order to get more bandwidth and meet the user demand. For instance two mechanisms are available for increasing the bandwidth; at Data link Layer Port Aggregation and at network Layer Redundancy Protocols. For both the solutions Standard protocols are available however both the solutions have certain limitations.

In the next section the evaluation process and limitations of these solutions are discussed in detail. After that new concept of Redundant Links Aggregation is discussed in detail with its performance comparison with standard Virtual Router Redundancy Protocol (VRRP).

## 2. RELATED WORK

As mentioned earlier first ever effort made in 1988 as ISDN however ISDN was a full fledge system and was Wide Area Network Technology However in Local Area Network (LAN) and especiallyat the Access and Distribution layer of the Network design, first effort was made by KALPANA during 1990[2].KALPANA proposed and implementedthe concept of Link Aggregation which is then acquired by Cisco in 1994. Cisco implement this concept with the name of EtherChannel[3]. Cisco also develop Port Aggregation Protocol (PAgP) which is used to handle the EtherChannel.This concept was proposed for purely for Ethernet technology.

The standardization process of Link Aggregation start in November 1997[4]. In 2000, IEEE 802.3ad group releases first version of standard protocol for Link Aggregation called Link Aggregation Control Protocol (LACP) [5]. Before that most venders provide their proprietary solution for aggregation capability. The IEEE 802.3 ad standard is revisited in 2008 and according to the suggestions of David Law the Link Aggregation standard is formally transferred under the umbrella of 802.1 as its sub-standard.Its identity is also changed as 802.1 AX-2008 [6][7]. IEEE describe Link aggregation as the method of using multiple Ethernet links or ports in parallel in order to increase the link bandwidth beyond the limits of any single link or port in Ethernet Technology[8].

Link Aggregation gives solutions for two problems related to Ethernet LAN Technology; Bandwidth limitation and redundancy in connections[9]. As the Ethernet bandwidths increases ten times by each previous bandwidth, so increase bandwidth upgrading of the equipment. The upgrading of equipment also require new media technology which consequently increase the budget. To solve this problem Link Aggregation is cost effective and appropriate solution. According to Link Aggregation Protocol user can bundle two or four or six or eight links together use it as single logical link[9]. However the bigger problems related to this solution are its technology dependency (as this solution was given for Ethernet) and Strict rules of combination. The rules LACP Link Aggregation includes: the number of participant links must even in number, the participant link must be of same type and bandwidth and number of link must not be exceeded from sixteen[10].
At Network layer another solutionfor increasing of bandwidth is available in form of Redundancy protocols. Redundancy protocols are used in those scenarios where more than one gateways (Internet Uplinks) are available. These protocols takes primary and backup gateways from the user and use primary link for sending all traffic while the backup links remain inactive. If the primary link goes down by any mean, the backup link becomes active[9].

Cisco systems is the pioneer of Redundancy protocol. In 1998 Cisco release Hot Standby Router Protocol (HSRP)their firstproprietary Redundancy Protocol[11].In the same year IETF release RFC 2338 which define Virtual Router Redundancy Protocol (VRRP) a proposed standard [12]. Then IETF release draft standard of VRRPduring 2004 in their RFC 3768[13] and in 2010 IETF release RFC 5798 which is current standard of Redundancy protocol[14].

Like Link Aggregation, Redundancy protocols do not have strict rules like even number of links, same type and bandwidth of links, however it do not combine links. Theytreat all the links as separate links. The issue related with this type of protocols are their inactive secondary links. Their primary links are always active while secondary links remain inactive and provide standby facility to the network. Cisco provide load balancing facility in Gateway Load Balancing Protocol (GLBP) their new proprietary protocol.This protocol provide three types of load balancing Host-Dependent, Round Robin and Weighted[15].In this competition GLBP fails due to its property of proprietary standard.

In short a mechanism is needed to take benefit from the redundant active links with no restriction of even number of links, technologicaldependency,participants with same propertiesand the solution must be cost effective. Furthermore that mechanism must also handle the gateways according to the cost in case of volume based charges.

## 3. CONCEPTUAL FRAMEWORK OF REDUNDANT LINKS AGGREGATION

In this research new concept of Redundant LinksAggregation(RLA) is proposed.RLA will provide luxury to the user to Aggregate or bundle links of different data carrier technologies together with no restrictions of even number of participant. These technologies can be of different nature including Local and Wide Area Network Technologies.It will alsofacilitate theuser to aggregate links with different bandwidth and forward the traffic according to the cost of link.
The Redundant Links Aggregation (RLA)system is divided into two core modules,the Link Control Module (LCM) and the Data Forwarding Module (DFM). Link Control Module (LCM) will be used to handle the links while Data ForwardingModule will be responsible forwarding the outgoing trafficusing different techniques.

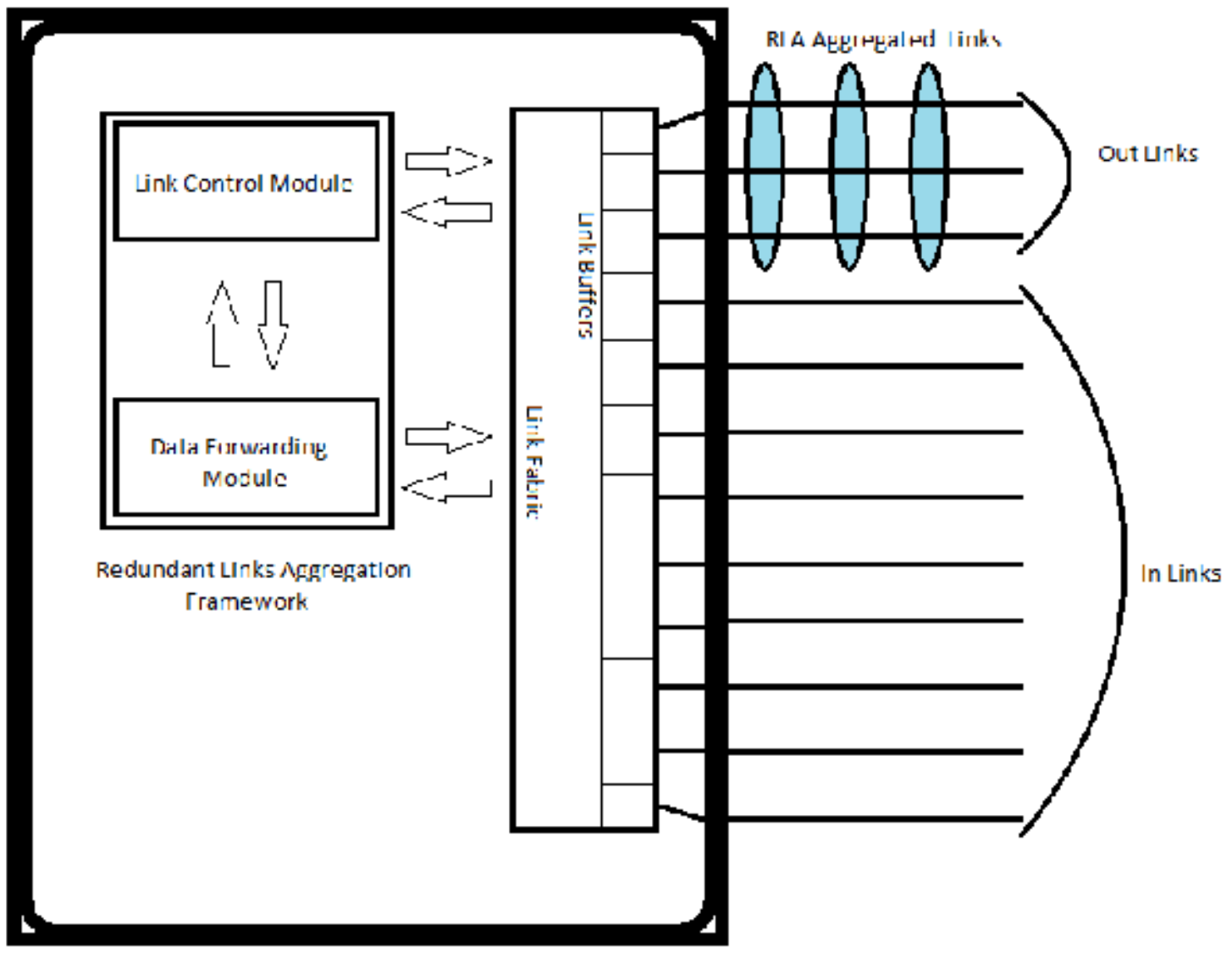

Figure 1. Framework of Redundant Link Aggregation

## 3.1 Link Control Module

The LCM will take aggregation parameters from the user. The aggregation parameters include Aggregation Group, Priority of the Link in a Groupand Cost of the Link. Aggregation Group parameter will be necessary, this will help the device to maintain multiple groups on a single device.Priority of the link will facilitate the user to set primary, secondary or tertiary link of the bundle.The cost parameter will be used in the Weighted Fair Queuing (WFQ) technique.

## 3.2Data Forwarding Module

After setting parametersof the aggregated bundle Data Forwarding Module will be used to forward the traffic over the links. This module will provide different techniques traffic forwarding including Round Robin, Weighted Fair Queue (WFQ) and Odd Load Balance.The user will get the freedom of choosing the forwarding mechanismaccording to the need.

In Round Robin fashion all the incoming traffic will be distributed across all the links in the bundle.This mechanism is good for the load distribution between the nodes having same bandwidth and type. However it also introduces the problem of out of order PDU (Protocol Data Unit) at the receiving system. This mechanism is good for short lived connections. However for long lived connections like Stream TCP connections and Multicast type of environment this mechanism is not suitable due to out of order PDU transmission[16].In Weighted Fair Queue (WFQ) technique the data will be distributed across all the links in the bundle according to the cost of the link.

The recommended method for the Redundant Links Aggregation is Odd Load Balance (OLB). In OLB technique, primary link will be active for any type of traffic. Once the traffic exceeds from threshold value of the buffer of the link, the secondary link will become active to aid primary link.If the traffic also exceeds from the threshold value of secondary link, next lower priority link will become active to aid previous two links. The process of deactivation of link is same like the process of activation of link in reverse order. When the traffic become low to the extent that high priority links can handle them then the low priority link will become deactivated.

### 3.3 Suggested Algorithm for Odd Load Balancing

The process of activation of link means filling of the link buffer with forwarding data. Although the links are up continuously but their buffer will be filled by forwarding data according to the load and weight of the link defined by the user.

The suggested algorithm for Odd load balancing is as follows:

```
While (Continue)
   {
      z = 0;
for ( i = 1;  i ≤ n; i + +)
         {
            z = z + + ;
if ( B [z]  ≤ T[z])
            Break ;
            }
      Send data to B [z]
   }
```

There are three variables “z”, “n” and “i" are used in the algorithm. “z” is used to access the array contents. “n” shows total number of links in a group, and “i” is used in a loop. Two Arrays “B[z]” and “T[z]” are also used which contains the value of current buffer size of all participated links and the threshold values of the buffers of all the participated links respectively.

The algorithm starts from continues WHILE loop, then z variable is initialized with zero. After that there is an FOR loop which checks wither i (the next variable) exceeds number of link or not. If the condition is true it will go further. In the body of the FOR loop there is an increment statement which will increment z, then the buffer of z link will check against the threshold value of z link. If condition is true means the buffer is not full so the loop will break and the data will be sent to buffer z. if the condition is not true the FOR loop will continue and will compare next link buffer with its own threshold. The flow of this algorithm is shown in figure 2.

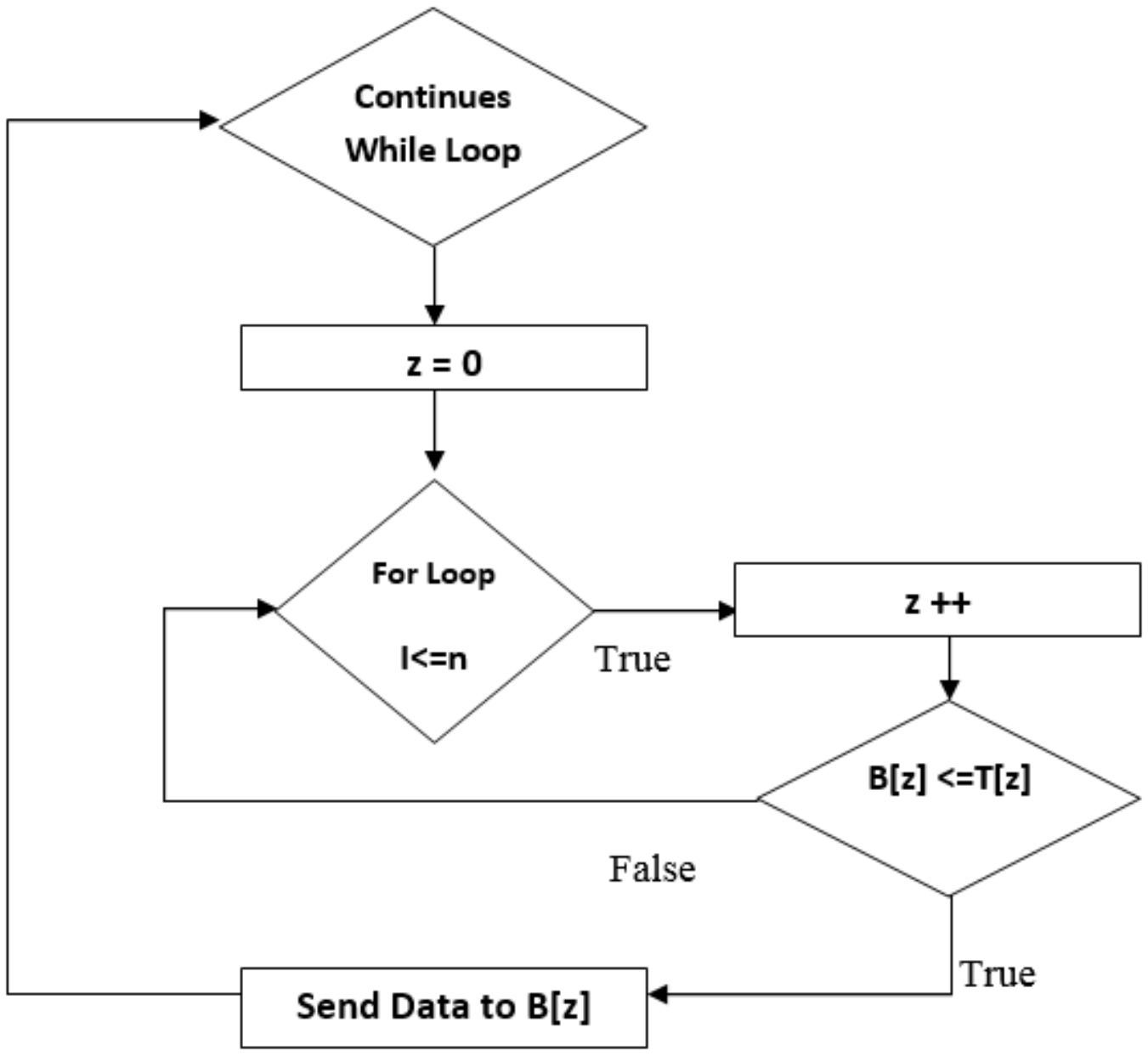

Figure2. Flow Chart of Odd Load Balancing Algorithm.

## 4. SIMULATION RESULTS

To evaluate the performance, Redundant Links Aggregation (RLA) is compared with Virtual Router Redundancy Protocol (VRRP). Two scenarios from the real world have been selected and both the mechanisms are tested against the demand of the network.

In first scenario 24 hours internet traffic of a university is taken. The available number of Internet connections is two i.e. 64Mbps and 32Mbps. The link of 64Mbps is used as primary link while link having 32Mbps will be used as backup link. Odd Load Balancing mechanism is used in Redundant Links Aggregation scheme.

In second scenario 24 hours internet traffic of another university is taken. The number of available connections in second scenario is three which includes two connections of 16Mbps and one connection of 4Mbps. 4Mbpsconnectionsis taken as primary connection according to the cost, while both16 Mbps connections are taken as secondary and tertiary. Odd Load Balancing mechanism is also taken for Redundant Links Aggregation scheme.

In both the graphs the demanded bandwidth is plotted against the supplied bandwidth. Orange line in the graph shows the demanded bandwidth of the network while the light blue area and green line shows theshows the supplied bandwidth of VRRP and RLA respectively.

### 4.1 Scenario 1

The result of first scenario shows that VRRP failed to satisfy demand of the network at peak hours. 1000 hours to 1600 hoursis the peak time of the network. At that time the demanded bandwidth is higher than 64Mbps while VRRP provides 64Mbps. However if we the RLA reaction in peak hours, its supply varies according to the demand. The maximum supplied bandwidth of RLA in that case is 96Mbps. As shown in figure 3.

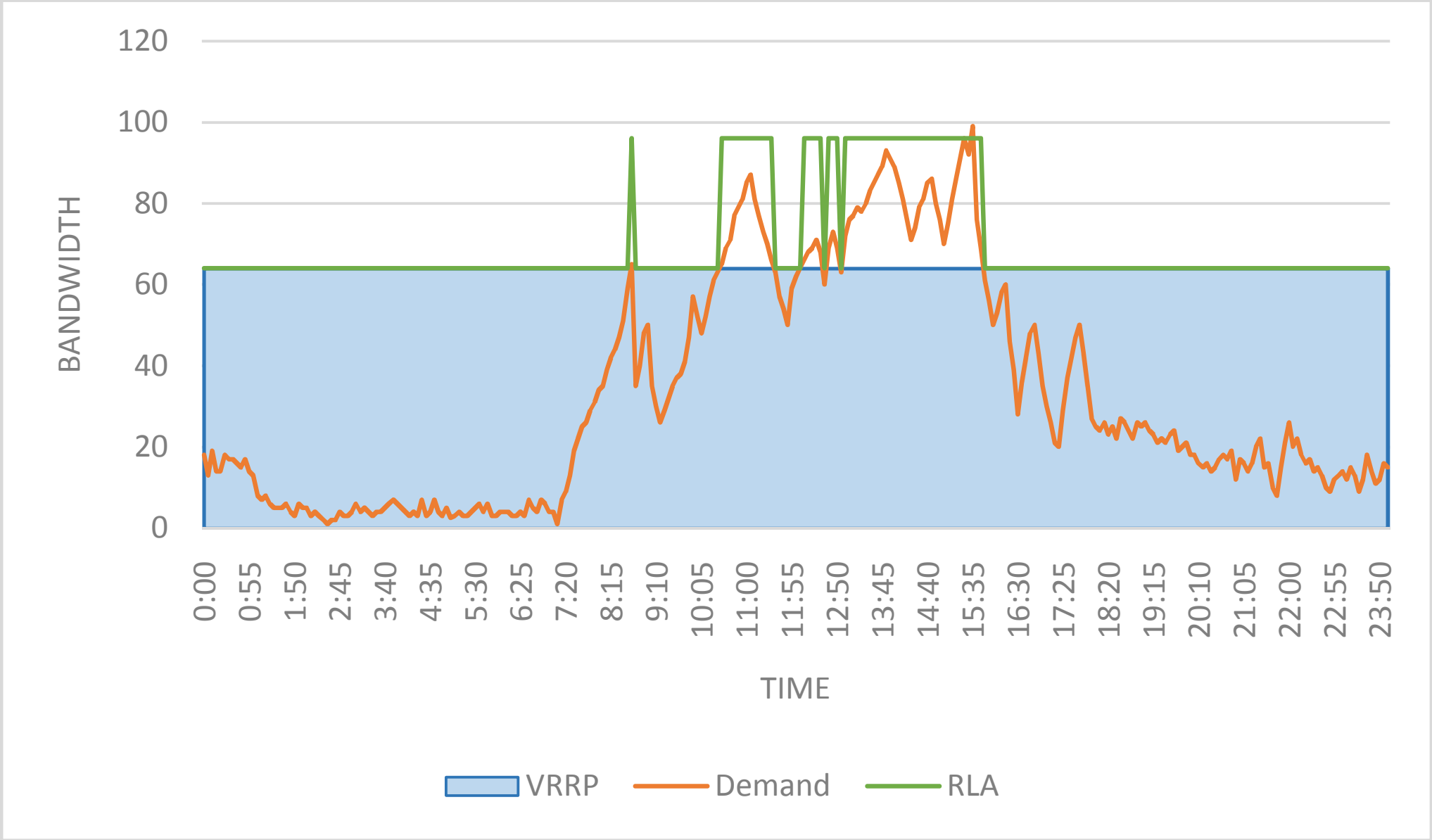

Figure 3. Scenario 1

## 4.2 Scenario 2

The second scenario tested in terms of cost effectiveness.The peak time of this network is between 1030 hours to 1600 hours.In peak hours RLA works in three levels. When demandexceeded from4Mbps, RLA open the secondary pathto aid the primary link. At 1000 hours and1600 hours when demand exceededfrom 20Mbps, RLA open the tertiary path as well which gives maximum bandwidth 36Mbps. However throughout the simulation VRRP provided a constant bandwidth of 16Mbps.

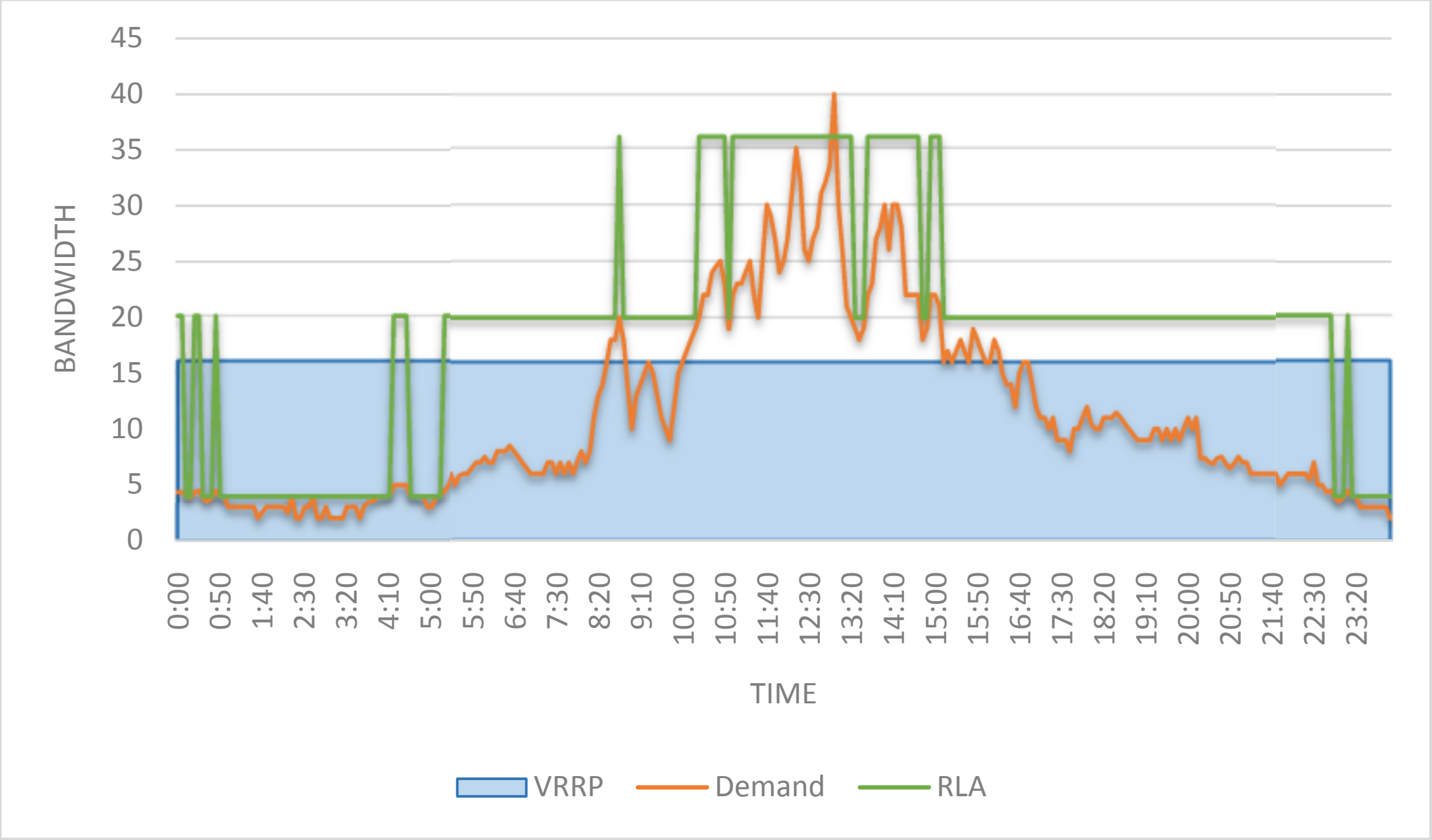

Graph. 2 Scenario 2

## 5. CONCLUSIONS

In overall, the outcome of this research came to prove that the new suggested Redundant Link Aggregation (RLA) scheme performance in the peak hours is batter then standard router redundancy protocol. Iteffectively make use of backup link which are remain idle while using the standard router redundancy protocol.The feature of adjusting the bandwidth according to the demand of the network makes it more effective in terms of cost. The Odd Load Balancing mechanism can be further configured according to the cost of the network. More economical connectionas primary link while less economical as secondary link. Which help the organization to save a notable sum of money per annum.

The work over Quality of Service of Redundant Link Aggregation should also be carried out in order to improve Quality of Service (QoS) for special protocols and services like Voice over IP, Video over IP, Multicast streaming Data and other Real time protocols.

## REFERENCES

[1] M. Decina and E. Scace, "CCITT Recommendations on the ISDN: A Review," Selected Areas in Communications, IEEE Journal on, vol. 4, pp. 320-325, 1986.
[2] C. Balakrishnan and M. Manikandan, "Link Aggregation Models and Services," International Journal of Science, Engineering and Technology Research, vol. 2, pp. pp: 038-042, 2013.
[3] O. Spatscheck, "Layers of Success," Internet Computing, IEEE, vol. 17, pp. 3-6, 2013.
[4] R. M. Wyatt, "Link aggregation," Google Patents, 2004.
[5] T. Wegmann and M. Gilmore, "IEEE 802.3 Ethernet Working Group," 2007.
[6] T. Kojima, M. Kanada, N. Inomata, and K. Kikuchi, "Communication apparatus, communication method, and computer program for LACP," EP Patent 2,362,588, 2011.
[7] D. Law, "IEEE 802.3 Maintenance," PDF, 2006.

[8] S.-M. Cho, "Ethernet communication apparatus, bridge thereof and connection device," U.S. Patent Application 10/339,566, 2003.

[9] D. Hucaby, CCNP BCMSN exam certification guide: CCNP self-study: Cisco Systems, 2004.

[10] D. Hucaby, CCNP switch 642-813 official certification guide: Cisco Press, 2010.

[11] D. Li, P. Morton, T. Li, and B. Cole, "Cisco hot standby router protocol (HSRP)," 1998.

[12] S. Knight, D. Weaver, D. Whipple, R. Hinden, D. Mitzel, P. Hunt, P. Higginson, M. Shand, and A. Lindem, "Virtual router redundancy protocol," RFC2338, April, 1998.

[13] R. Hinden, "Virtual router redundancy protocol (VRRP)," 2004.

[14] S. Nadas, "Virtual Router Redundancy Protocol (VRRP) Version 3 for IPv4 and IPv6," 2010.

[15] T. J. Nosella and I. H. Wilson, "Gateway load balancing protocol," Google Patents, 2011.

[16] C. Klonowski, H. J. Seeger, J. Kim, S. G. Bueno, and K. W. Jeong, AIX Version 4.3 to 5L Migration Guide: IBM, International Technical Support Organization, 2003.

## AUTHORS

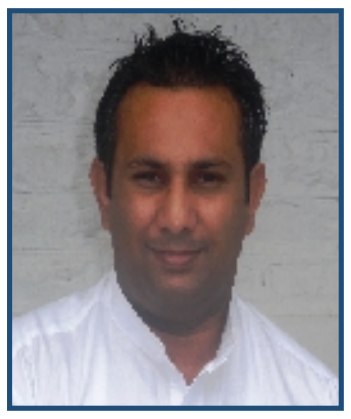

Mr Rafiullah Khan, Lecturer Institute of Business and Management Sciences (IBMS), The University of Agriculture Peshawar Pakistan received his BS (Hons) degree from University of Peshawar in 2007 and MS degree in Information Technology from the Institute of Management Science (IM|Sciences) Peshawar, Pakistan in 2010. Currently, he is perusing Ph.D program in the Department of Computer Science of the University of Peshawar. His fields of specialization are Semantic Web and Computer Based Communication Networks.

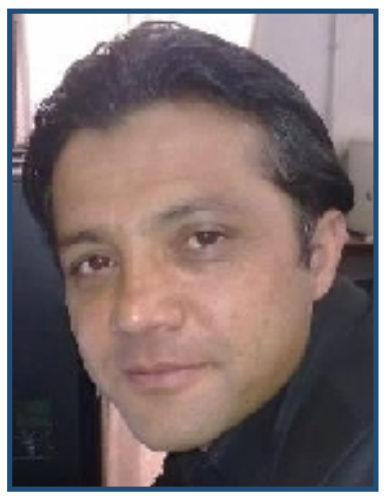

Mr Shaukat Ali, Assistant Professor, FG Degree College Peshawar Cantt, KPK, Pakistanreceived M.Sc degree in computer science from University of Peshawar, KPK, Pakistan in 2002, and the MS degree in Information Technology from the same university in 2007. Currently, he is perusing Ph.D program in the Department of Computer Science of the above mentioned university. His fields of specialization are Semantic Web and Computer Based Communication Networks. He has authored 5 research papers, published in internationalJournals and presented in international conferences.